\documentclass[1p]{elsarticle}

\usepackage{hyperref}
\usepackage{booktabs}
\usepackage{adjustbox}
\usepackage{xcolor}
\usepackage{ulem}
\usepackage{soul}

\usepackage[symbol]{footmisc}
\makeatletter
\def\ps@pprintTitle{%
   \let\@oddhead\@empty
   \let\@evenhead\@empty
   \def\@oddfoot{\reset@font\hfil\thepage\hfil}
   \let\@evenfoot\@oddfoot
}
\makeatother
\journal{Helyion}

\begin{document}
\begin{frontmatter}


\title{Hippocampus Segmentation on Epilepsy and Alzheimer's Disease Studies with Multiple Convolutional Neural Networks\footnote[2]{Source code: \url{https://github.com/MICLab-Unicamp/e2dhipseg}, published in Heliyon \url{https://www.sciencedirect.com/science/article/pii/S2405844021003315} (DOI: 10.1016/j.heliyon.2021.e06226).}}

\author[mainaddress]{Diedre Carmo}
\author[secaddress]{Bruna Silva}
\author{Alzheimer's Disease Neuroimaging Initiative\footnote[1]{Part of the data used in preparation of this article were obtained from the Alzheimer's Disease Neuroimaging Initiative (ADNI) database (\url{adni.loni.usc.edu}). As such, the investigators within the ADNI contributed to the design and implementation of ADNI and/or provided data but did not participate in analysis or writing of this report. A complete listing of ADNI investigators can be found at: \url{http://adni.loni.usc.edu/wp-content/uploads/how_to_apply/ADNI_Acknowledgement_List.pdf}}}
\author[secaddress]{Clarissa Yasuda}
\author[mainaddress]{Letícia Rittner}
\author[mainaddress]{Roberto Lotufo}

\address[mainaddress]{School of Electrical and Computer Engineering, UNICAMP, Campinas, São Paulo, Brazil}

\address[secaddress]{Faculty of Medical Sciences, UNICAMP, Campinas, São Paulo, Brazil}

\begin{abstract}
\textbf{Background:} Hippocampus segmentation on magnetic resonance imaging is of key importance for the diagnosis, treatment decision and investigation of neuropsychiatric disorders. Automatic segmentation is an active research field, with many recent models using deep learning. Most current state-of-the art hippocampus segmentation methods train their methods on healthy or Alzheimer's disease patients from public datasets. This raises the question whether these methods are capable of recognizing the hippocampus on a  different domain, that of epilepsy patients with hippocampus resection.\\
\textbf{New Method: } In this paper we present a state-of-the-art, open source, ready-to-use, deep learning based hippocampus segmentation method. It uses an extended 2D multi-orientation approach, with automatic pre-processing and orientation alignment. The methodology was developed and validated using HarP, a public Alzheimer's disease hippocampus segmentation dataset.\\ 
\textbf{Results and Comparisons: } We test this methodology alongside other recent deep learning methods, in two domains: The HarP test set and an in-house epilepsy dataset, containing hippocampus resections, named HCUnicamp. We show that our method, while trained only in HarP, surpasses others from the literature in both the HarP test set and HCUnicamp in Dice. Additionally, Results from training and testing in HCUnicamp volumes are also reported separately, alongside comparisons between training and testing in epilepsy and Alzheimer's data and vice versa.\\ 
\textbf{Conclusion: } Although current state-of-the-art methods, including our own, achieve upwards of 0.9 Dice in HarP, all tested methods, including our own, produced false positives in HCUnicamp resection regions, showing that there is still room for improvement for hippocampus segmentation methods when resection is involved.

\end{abstract}


\end{frontmatter}

\section{Introduction}

The hippocampus is a small, medial, subcortical brain structure related to long and short term memory~\cite{andersen2007hippocampus}. The hippocampus can be affected in shape and volume by different pathologies, such as the neurodegeneration associated to Alzheimer's disease~\cite{petersen2010alzheimer}, or surgical intervention to treat temporal lobe epilepsy~\cite{ghizoni2017clinical}. Hippocampal segmentation from magnetic resonance imaging (MRI) is of great importance for research of neuropsychiatric disorders and can also be used in the preoperatory investigation of pharmacoresistant temporal lobe epilpesy~\cite{ghizoni2015modified}.  The medical research of these disorders usually involves manual segmentation of the hippocampus, requiring time and expertise in the field. The high-cost associated to manual segmentation has stimulated the search for effective automatic segmentation methods. Some of those methods, such as FreeSurfer~\cite{fischl2012freesurfer}, are already used as a starting point for a manual finer segmentation later~\cite{mccarthy2015comparison}.

While conducting research on epilepsy and methods for hippocampus segmentation, two things raised our attention. Firstly, the use of deep learning and Convolutional Neural Networks (CNN) is in the spotlight. with most of the recent hippocampus segmentation methods featuring them. Secondly, many of these methods rely on publicly available datasets for training and evaluating and therefore have access only to healthy scans, or patients with Alzheimer's disease. This raises the concern that automated methods might only be prepared to deal with features present in the public Alzheimer's and healthy subjects datasets, such as ADNI and the Multi Atlas Labeling Challenge (MALC).

Considering these facts, we present an improved version of our own deep learning based hippocampus segmentation method~\cite{carmo2019midl}, compared with other recent methods~\cite{roy2019quicknat, thyreau2018segmentation, isensee2017brain}. We use the public Alzheimer's HarP dataset for training and initial testing comparisons with other methods. As an additional test dataset, an in-house epilepsy dataset named HCUnicamp is used. It contains scans from patients with epilepsy (pre and post surgical removal of hippocampus), with  different patterns of atrophy compared to that observed both in the Alzheimer's data and healthy subjects. It is important to note that HCUnicamp is not involved in our method's training or methodological choices, to allow for fair comparisons with other methods. Without comparing to other methods, we also report results of involving HCUnicamp epilepsy volumes in training.

\subsection{Contributions}

In summary, the main contributions of this paper are as follows:

\begin{itemize}
\item A readily available hippocampus segmentation methodology under the MIT license, consisting of an ensemble of 2D CNNs coupled with traditional 3D post processing, achieving state of the art performance in HarP public data, and using recent advancements from the deep learning literature.

\item An evaluation of recent hippocampus segmentation methods in our epilepsy test dataset, HCUnicamp, that includes post-operatory images of patients without one of the hippocampi. In this evaluation, our method is only trained in public HarP volumes, therefore our methodology has no bias related to this task. We show that our method is also superior in this domain, although no method was able to achieve more than 0.8 Dice in this dataset, according to our manual annotations. As far as we know, that has not been explored before with recent Deep Learning methods.

\item A final experiment includes epilepsy HCUnicamp volumes in training, without changing the methodology and with no comparisons to other methods, which resulted in better performance on epilepsy cases. The effects of mixing data from both datasets in training are explored.

\end{itemize}

This paper is organized as follows: Section~\ref{sec:lit} presents a literature review of recent deep learning based hippocampus segmentation methods. Section~\ref{sec:data} introduces more details to the two datasets involved in this research. A detailed description of our hippocampus segmentation methodology is in Section~\ref{sec:methodology}. Section~\ref{sec:results} has experimental results from our methodology development, qualitative and quantitative comparisons with other methods in HarP and HCUnicamp, and results of involving HCUnicamp volumes in traning. Sections~\ref{sec:discussion} and~\ref{sec:conclusion} have, respectively, extended discussion of those results and conclusion. More details to the training and hyperparameter optimization process are in the appendix.

\section{Hippocampus Segmentation with Deep Learning}
\label{sec:lit}
Before the rise of deep learning methods in medical imaging segmentation, most hippocampus segmentation methods used some form of optimization of registration and deformation to atlas(es)~\cite{wang2013multi, iglesias2015multi, pipitone2014multi, fischl2012freesurfer, chincarini2016integrating, platero2017combining}. Even today, medical research uses results from FreeSurfer~\cite{fischl2012freesurfer}, a high impact multiple brain structures segmentation work, available as a software suite. Those atlas-based methods can produce high quality segmentations, taking, however, around 8 hours in a single volume. Lately, a more time efficient approach appeared in the literature, namely the use of such atlases as training volumes for CNNs. Deep learning methods can achieve similar overlap metrics while predicting results in a matter of seconds per volume~\cite{chen2017hippocampus, xie2018near, wachinger2018deepnat, thyreau2018segmentation, roy2019quicknat, ataloglou2019fast, dinsdale2019spatial}. 

Recent literature on hippocampus segmentation with deep learning is exploring different architectures, loss functions and overall methodologies for the task. One approach that seems to be common to most of the studies involves the combination of 2D or 3D CNNs, and patches as inputs in the training phase. Note that some works focus on hippocampus segmentation, while some attempt segmentation of multiple neuroanatomy. Following, a brief summary of each of those works.

Chen et al.~\cite{chen2017hippocampus} reports 0.9 Dice~\cite{sudre2017generalised} in 10-fold 110 ADNI~\cite{petersen2010alzheimer} volumes with a novel CNN input idea. Instead of using only the triplanes as patches, it also cuts the volume in six more diagonal orientations. This results in 9 planes, that are fed to 9 small modified U-Net~\cite{ronneberger2015u} CNNs. The ensemble of these U-Nets constructs the final result. 

Xie et al.~\cite{xie2018near} trains a voxel-wise classification method using triplanar patches crossing the target voxel. They merge features from all patches into a Deep Neural Network with a fully connected classifier alongside standard use of ReLU activations and softmax~\cite{krizhevsky2012imagenet}. The training patches come only from the approximate central area the hippocampus usually is, balancing labels for 1:1 foreground and background target voxels. Voxel classification methods tend to be faster than multi-atlas methods, but still slower than Fully Convolutional Neural Networks. 

DeepNat from Wachinger et al.~\cite{wachinger2018deepnat} achieves segmentation of 25 structures with a 3D CNN architecture. With a hierarchical approach, a 3D CNN separates foreground from background and another 3D CNN segments the 25 sub-cortical structures on the foreground. Alongside a proposal of a novel parametrization method replacing coordinate augmentation, DeepNat uses 3D Conditional Random Fields as post-processing. The architecture is a voxelwise classification, taking into account the classification of neighbor voxels. This work's results mainly focuses on the MALC dataset, with around 0.86 Dice in hippocampus segmentation.  

Thyreau et al.~\cite{thyreau2018segmentation}'s model, named Hippodeep, uses CNNs trained in a region of interest (ROI). However, where we apply one CNN for each plane of view, Thyreau et al. uses a single CNN, starting with a planar analysis followed by layers of 3D convolutions and shortcut connections. This study used more than 2000 patients, augmented to around 10000 volumes with augmentation. Initially the model is trained with FreeSurfer segmentations, and later fine-tuned using volumes which the author had access to manual segmentations, the gold standard. Thyreau's method requires MNI152 registration of input data, which adds around a minute of computation time, but the model is generally faster than multi-atlas or voxel-wise classification, achieving generalization in different datasets, as verified by Nogovitsyn et al.~\cite{nogovitsyn2019testing}.

QuickNat from Roy et al.~\cite{roy2019quicknat} achieves faster segmentations than DeepNat by using a multiple CNN approach instead of voxel-wise classification. Its methodology follows a consensus of multiple 2D U-Net like architectures specialized in each slice orientation. The use of FreeSurfer~\cite{fischl2012freesurfer} masks over hundreds of public data to generate silver standard annotations allows for much more data than usually available for medical imaging. Later, after the network already knows to localize the structures, it is finetuned to more precise gold standard labels. Inputs for this method need to conform to the FreeSurfer format. 

Ataloglou et al.~\cite{ataloglou2019fast} recently displayed another case of fusion of multiple CNN outputs, specialized into axial, coronal and sagittal orientations, into a final hippocampus segmentation. They used U-Net like CNNs specialized in each orientation, followed by error correction CNNs, and a final average fusion of the results. They went against a common approach in training U-Nets of using patches during data augmentation, instead using cropped slices. This raises concerns about overfitting to the used dataset, HarP~\cite{boccardi2015training}, supported by the need of finetuning to generalize to a different dataset.

Dinsdale et al.~\cite{dinsdale2019spatial} mixes knowledge from multi-atlas works with deep learning, by using a 3D U-Net CNN to predict a deformation field from an initial binary sphere to the segmentation of the hippocampus, achieving around 0.86 DICE on Harp. Interestingly, trying an auxiliary classification task did not improve segmentation results. 

It is known that deep learning approaches require a relatively large amount of varied training data. Commonly used forms of increasing the quantity of data in the literature include using 2D CNNs over regions (patches) of slices, with some form of patch selection strategy. The Fully Convolutional Neural Network (FCNN) U-Net~\cite{ronneberger2015u} architecture has shown potential to learn from relatively small amounts of data with their decoding, encoding and concatenation schemes, even working when used with 3D convolutions directly in a 3D volume~\cite{isensee2017brain}. 

Looking at these recent works, one can confirm the segmentation potential of the U-Net architecture, including the idea of an ensemble of 2D U-Nets instead of using a single 3D one, as we~\cite{carmo2019extended, carmo2019midl}, some simultaneous recent work~\cite{roy2019quicknat, ataloglou2019fast}, or even works in other segmentation problems~\cite{lucena2018silver} presented. In this paper, some of those methods were reproduced for comparison purposes in our in-house dataset, namely~\cite{roy2019quicknat, thyreau2018segmentation}, including a 3D UNet architecture test from~\cite{isensee2017brain}.

As far as we know, there is no study applying recent deep learning methods trained on public data, such as HarP and MALC, to MRI scans of epilepsy including hippocampus resection cases. We also include, separately, an attempt to train on such data.

\section{Data}
\label{sec:data}
\begin{figure}[htbp]
\begin{center}
\begin{minipage}[b]{.5\textwidth}
  \includegraphics[width=\textwidth]{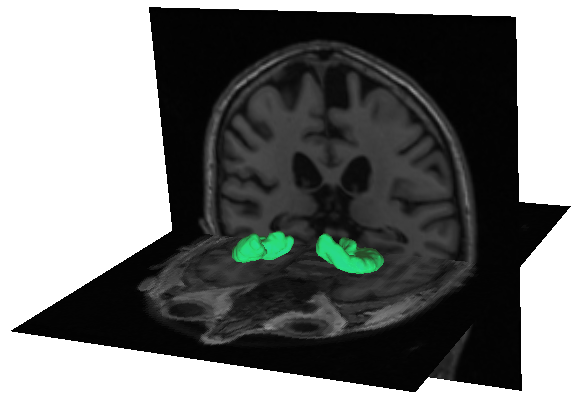}
  \centerline{(a)}\medskip
\end{minipage}
\begin{minipage}[b]{.3\textwidth}
  \includegraphics[width=\textwidth]{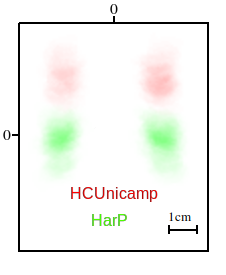}
  \centerline{(b)}\medskip
\end{minipage}
\begin{minipage}[b]{.3\textwidth}
  \includegraphics[width=\textwidth]{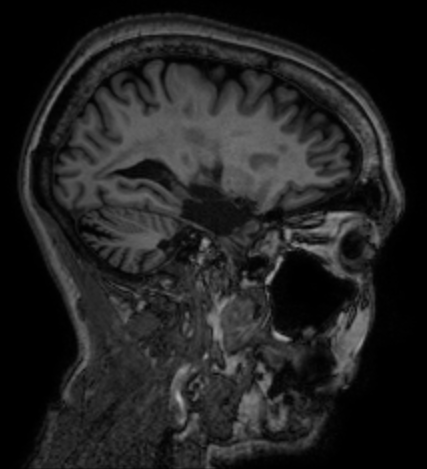}
  \centerline{(c)}\medskip
\end{minipage}
\begin{minipage}[b]{.23\textwidth}
  \includegraphics[width=\textwidth]{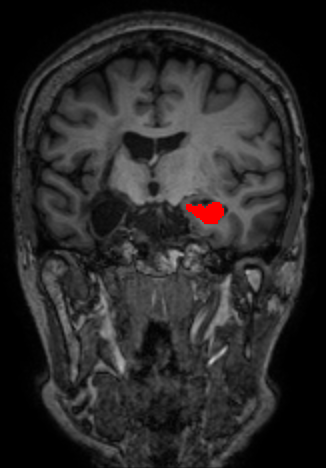}
  \centerline{(d)}\medskip
\end{minipage}
\begin{minipage}[b]{.257\textwidth}
  \includegraphics[width=\textwidth]{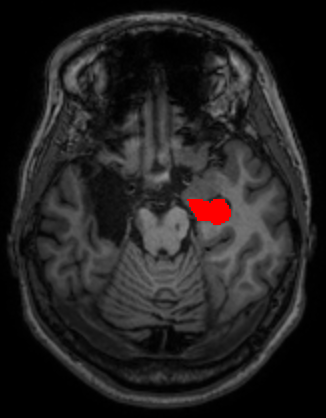}
  \centerline{(e)}\medskip
\end{minipage}
\caption{(a) 3D rendering of the manual annotation (in green) of one of the HarP dataset volumes. In (b), a coronal center crop slice of the average hippocampus mask for all volumes in HarP (green) and HCUnicamp (red), shows different head alignment. Zero corresponds to the center. (c) Sagittal, (d) Coronal and (e) Axial HCUnicamp slices from a post-operative scan with annotations in red.}
\label{fig:bigdata}
\end{center}
\end{figure}

This study uses mainly two different datasets: one collected locally for an epilepsy study, named HCUnicamp; and one public from the ADNI Alzheimer's study, HarP. HarP is commonly used in the literature as a hippocampus segmentation benchmark. The main difference between the datasets is, the lack of one of the hippocampi in 70\% of the scans from HCUnicamp, as these patients underwent surgical removal (Figure~\ref{fig:bigdata}). 

Although our method needs input data to be in the MNI152~\cite{brett2001using} orientation, data from those datasets are in native space and are not registered. We provide an automatic orientation correction by rigid registration as an option when predicting in external volumes, to avoid orientation mismatch problems.

\subsection{HarP}
This methodology was developed with training and validation on HarP~\cite{boccardi2015training}, a widely used benchmark dataset in the hippocampus segmentation literature. HarP uses data from the Alzheimer’s disease Neuroimaging Initiative (ADNI) database (adni.loni.usc.edu). The ADNI was launched in 2003 as a public-private
partnership, led by Principal Investigator Michael W. Weiner, MD. The primary goal of ADNI has been to test whether serial magnetic resonance imaging (MRI), positron emission tomography (PET), other biological markers, and clinical and neuropsychological assessment can be combined to measure the progression of mild cognitive impairment (MCI) and early Alzheimer’s disease (AD).

The full HarP release contains 135 T1-weighted MRI volumes. Alzheimer's disease classes are balanced with equal occurrence of control normal (CN), mild cognitive impairment (MCI) and alzheimer's disease (AD) cases~\cite{petersen2010alzheimer}. Volumes were minmax intensity normalized between 0 and 1, and no volumes were removed. Training with stratified hold-out was performed with 80\% training, 10\% validation and 10\% testing, while k-Folds, when used, consisted of 5 folds, with no overlap on the test sets.

\subsection{HCUnicamp}
HCUnicamp was collected inhouse, by personnel from the Brazilian Institute of Neuroscience and Neurotechnology (BRAINN) at UNICAMP's \textit{Hospital de Clínicas}. This dataset contains 190 T1-weighted 3T MRI acquisitions, in native space. 58 are controls and 132 are epilepsy patients. From those epilepsy images, 70\% had one of the hippocampus surgically removed, resulting in a very different shape and texture than what is commonly seen in public datasets (Figure~\ref{fig:bigdata}). More details about the surgical procedure can be found in~\cite{ghizoni2015modified, ghizoni2017clinical}. All volumes have manual annotations of the hippocampus, performed by one rater. The voxel intensity is minmax normalized, between 0 and 1, per volume. This data acquisition and use was approved by an Ethics and Research Committee (CEP/Conep, number 3435027).  

A comparison between the datasets can be seen in Figure~\ref{fig:bigdata}. The difference in mean mask position due to the inclusion of neck in HCUnicamp is notable, alongside with the lower presence of left hippocampus labels due to surgical intervention for epilepsy (Figure~\ref{fig:bigdata}b). 

To investigate the performance of different methods in terms of dealing with the absence of hippocampus and unusual textures, we used the whole HCUnicamp dataset (considered a different domain) as a final test set. Our methodology was only tested in this dataset at the end, alongside other methods. Results on HCUnicamp were not taken into consideration for our method's methodological choices, to allow for fair comparisons with other methods, treating this data as a true final test set.

A final additional experiment attempts to learn from the epilepsy data, dividing HCUnicamp in a balanced hold-out of 70\% training, 10\% validation and 20\% testing. These subsets are called HCU-Train, HCU-Validation and HCU-Test for clarity.

\section{Segmentation Methodology}
\label{sec:methodology}

\begin{figure}[ht]
\begin{center}
\includegraphics[width=\textwidth]{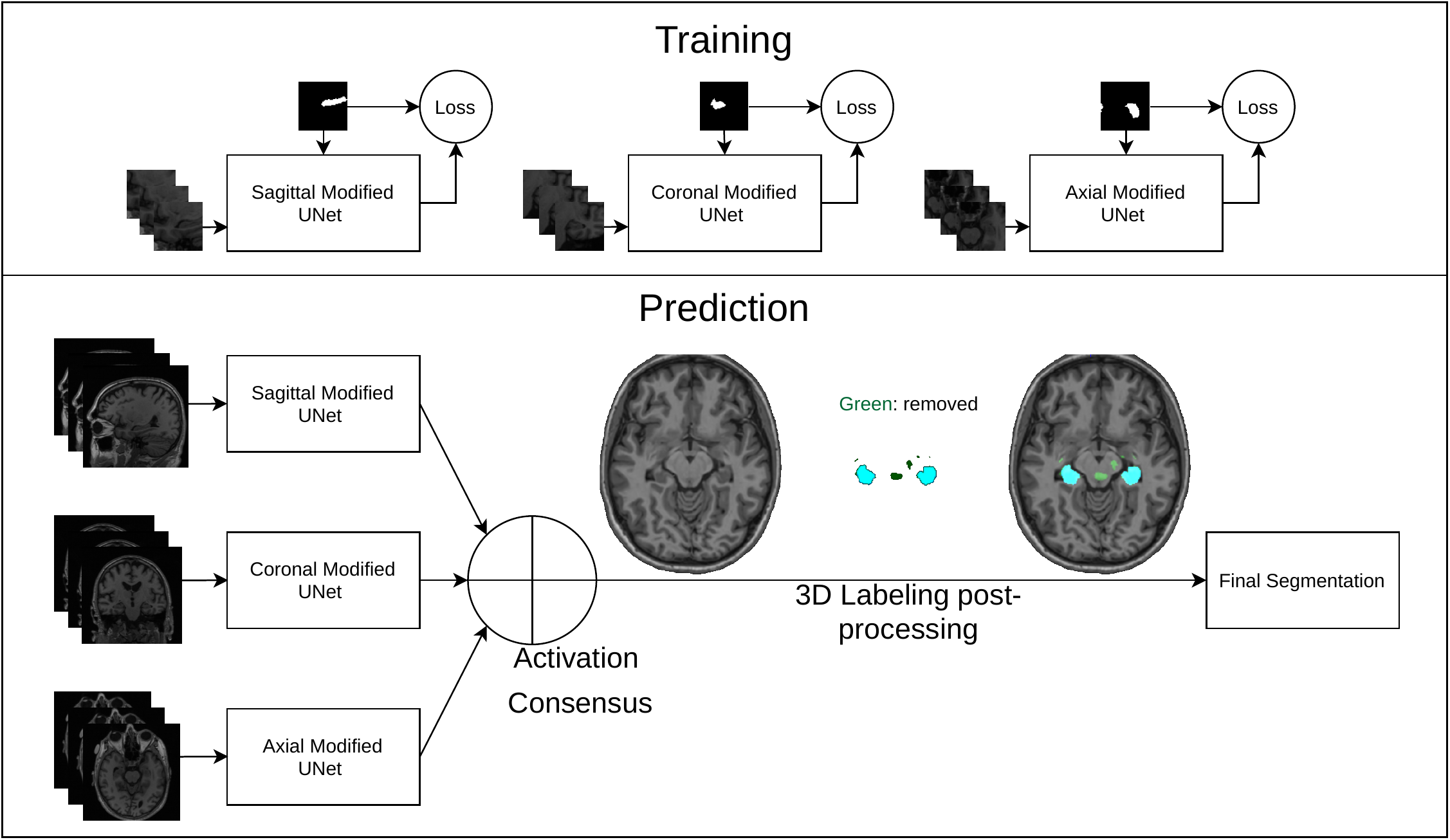}\\
\end{center}
\caption{\label{fig:outline} The final segmentation volume is generated by taking into account activations from three FCNNs specialized on each 2D orientation. Neighboring slices are taken into account in a multi-channel approach. Full slices are used in prediction time, but training uses patches.}
\end{figure}

In this section, the general methodology (Figure \ref{fig:outline}) for our hippocampus segmentation method is detailed. Three orientation specialized 2D U-Net CNNs are utilized, inspired by Lucena Et Al's work~{\cite{lucena2018silver}}. The activations from the CNNs are merged into an activation consensus. Each network's activations for a given input volume are built slice by slice. The three activation volumes are averaged into a consensus volume, which is post-processed into the final segmentation mask. 

\begin{figure}[ht]
\begin{center}
\includegraphics[width=\textwidth]{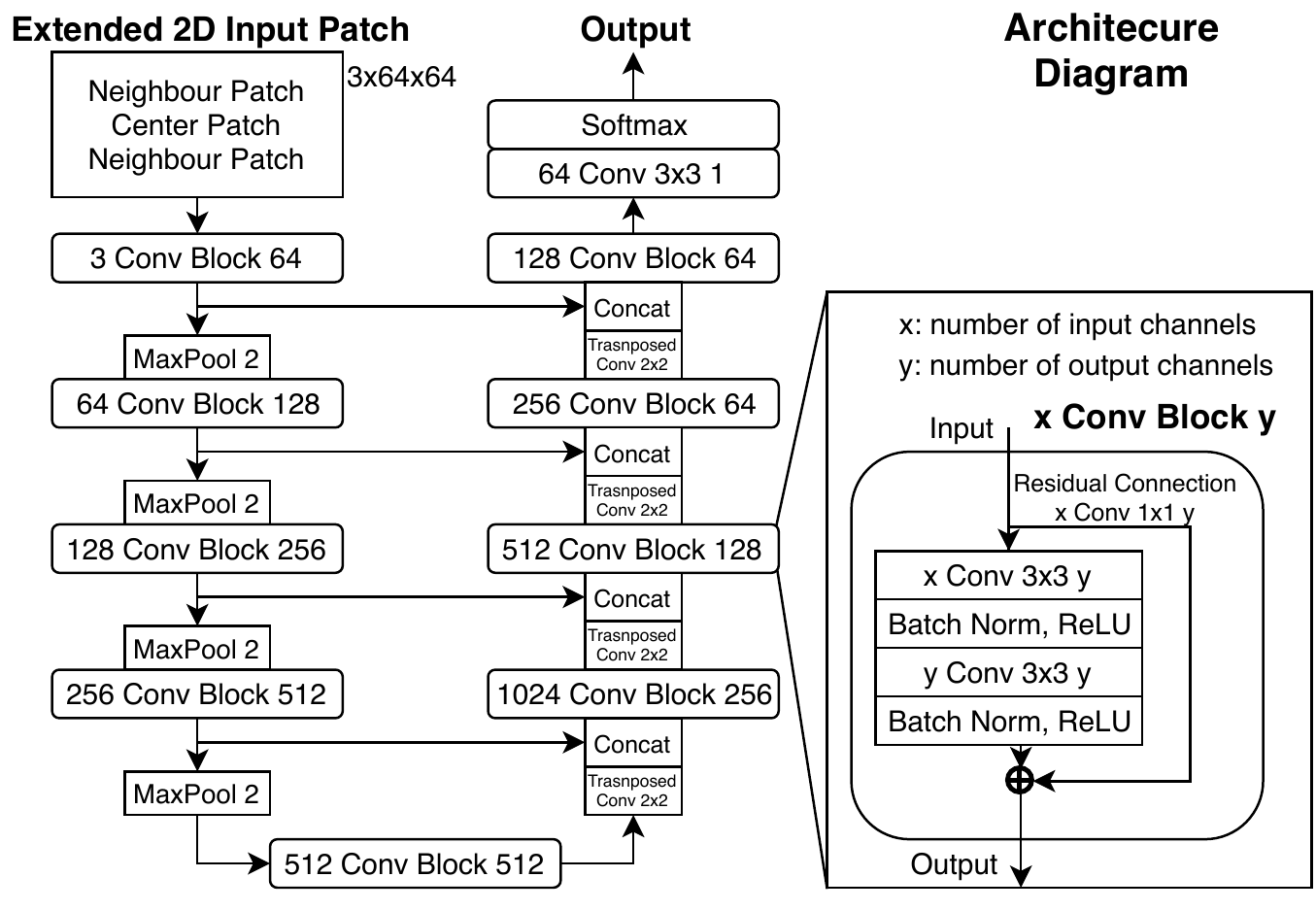}
\end{center}
\caption{\label{fig:arch} Final architecture of each modified U-Net in figure \ref{fig:outline}. Of note in comparison to the original U-Net is the use of BatchNorm, residual connections in each convolutional block, the 3 channel neighbour patches input and the sigmoid output limitation. Padding is also used after convolutions.}
\end{figure}

The basic structure of our networks is inspired by the U-Net FCNN architecture \cite{ronneberger2015u}. However, some modifications based on other successful works were applied to the architecture (Figure \ref{fig:arch}). Those modifications include: instead of one single 2D patch as input, two neighbour patches are concatenated leaving the patch corresponding to the target mask in the center~\cite{marianaextended2D}. Residual connections based on ResNet \cite{he2016deep} between the input and output of the double convolutional block were added, as 1x1 2D convolutions to account for different number of channels. Batch normalization was added to each convolution inside the convolutional block, to accelerate convergence and facilitate learning \cite{ioffe2015batch}. Also, all convolutions use padding to keep dimensions and have no bias. This works uses VGG11 \cite{simonyan2014very} weights in the encoder part of the U-Net architecture, as in \cite{iglovikov2018ternausnet}.

During prediction time, slices for each network are extracted with a center crop. When building the consensus activation volume, the resulting activation is padded back to the original size. For training, this method uses patches. Patches are randomly selected in runtime. Patches can achieve many possible sizes, as long as it accommodates the number of spatial resolution reductions present in the network.

A pre-defined percentage of the patches are selected from a random point of the brain, allowing for learning of what structures are not the hippocampus. Those are called negative patches. On the other hand, positive patches are always centered on a random point of the hippocampus border. In a similar approach to Pereira et al.~\cite{marianaextended2D}'s Extended 2D, adjacent patches (slices on evaluation) are included in the network's input as additional channels (Figure~\ref{fig:outline}). The intention is for the 2D network to take into consideration volumetric information adjacent to the region of interest, hence the name for the method, Extended 2D Consensus Hippocampus Segmentation (E2DHipseg). This approach is inspired by how physicians compare neighbor slices in multiview visualization when deciding if a voxel is part of the analyzed structure or not.

Data augmentation is used to improve our dataset variance and avoid overfitting. All augmentations perform a random small runtime modification to the data. Random augmentations include intensity modification ($[-0.05, 0.05]$), rotation and scale ($[-10, 10]$) and gaussian noise with $0$ mean and $0.0002$ variance.

\subsection{Loss Function}
Dice~\cite{sudre2017generalised} is an overlap metric widely used in the evaluation of segmentation applications. Performance in this paper is mainly evaluated with Dice, by comparisons with the manual gold standard. Dice can be defined as:

\begin{equation}
    \frac{2\sum_{i}^{N}p_{i}g_{i}}{\sum_{i}^{N}p_{i}^{2} + \sum_{i}^{N}g_{i}^{2}}
\end{equation}

Where the sums run over the N voxels, of the predicted binary segmentation
volume $p_{i} \in P$ and the ground truth binary volume $g_{i} \in G$. For conversion from a metric to a loss function, one can simply optimize $1 - Dice$, therefore optimizing a segmentation overlap metric. This is referred here as Dice Loss.

To take into account background information, a Softmax of two-channels representing background and foreground can be used as an output. In this case, Generalized Dice Loss (GDL)~\cite{sudre2017generalised} and Boundary Loss, a recent proposal of augmentation to GDL from Kervadec et al.~\cite{pmlr-v102-kervadec19a} were considered as loss options.

Generalized Dice Loss weights the loss value by the presence of a given label in the target, giving more importance to less present labels. This solves the a class imbalance problem that would emerge when using Dice Loss while including background as a class. 

Boundary Loss takes into consideration alongside the ``regional" loss (e.g. GDL), the distance between boundaries of the prediction and target, which does not gives any weight to the area of the segmentation. Kervadec's work suggests that a loss functions that takes into account boundary distance information can improve results, specially for unbalanced datasets. However, one needs to balance the contribution of both components with a weight, defined as $\alpha$ in the following Boundary Loss (B) equation:

\begin{equation}
B(p, g) = \alpha~G(p, g) + (1-\alpha)~S(p, g)
\end{equation}

Where G is GDL, regional component of the loss function, and S is the surface component, that operates on surface distances. The weight factor $\alpha$ changes from epoch to epoch. The weight given to the regional loss is shifted to the surface loss, with $\alpha$ varying from 1 in the first epoch to 0 in the last epoch. We followed the original implementation in~\cite{pmlr-v102-kervadec19a}.

\subsection{Consensus and Post-processing}
The consensus depicted in Figure \ref{fig:outline} consists of taking the average from the activations of all three CNNs. A more advanced approach of using a 4th, 3D, U-Net as the consensus generator was also attempted.

After construction of the consensus of activations, a threshold is needed to binarize the segmentation. We noticed that sometimes, small structures of the brain similar to the hippocampus could be classified as false positives. To remove those false positives, a 3D labeling implementation from \cite{dougherty2003hands} was used, with subsequent removal of small non-connected volumes, keeping the 2 largest volumes, or 1 if a second volume is not present (Figure \ref{fig:outline}). This post processing is performed after the average consensus of all networks and threshold application.

\section{Experiments and Results}
\label{sec:results}

This section presents quantitative and qualitative comparisons with other methods in HarP and HCUnicamp. The appendix showcases more detailed experiments on the segmentation methodology, displaying differences in Dice in the HarP test set, resulting from our methodological choices.


\subsection{Quantitative Results}
In this section, we report quantitative results of our method and others from the literature in both HarP and HCUnicamp. For comparison's sake, we also trained an off-the-shelf 3D U-Net architecture, from Isensee et al.~\cite{isensee2017brain}, originally a Brain Tumor segmentation work. Isensee's architecture was trained with ADAM and HarP 3D center crops as input.  

For the evaluation with the QuickNat~\cite{roy2019quicknat} method, volumes and targets needed to be conformed to its required format, causing interpolation. As far as we know, the method does not have a way to return its predictions on the volume's original space. DICE was calculated with the masks on the conformed space. Note that QuickNat performs segmentation of multiple brain structures.

\subsubsection{HarP}

The best hold-out mean Dice is $0.9133$. In regards to specific Alzheimer's classes in the test set, our method achieves $0.9094$ Dice for CN, $0.9378$ for MCI and $0.9359$ for AD cases. When using a hold-out approach in a relatively small dataset such as HarP, the model can be overfitted to better results in that specific test set. With that in mind, we also report results with cross validation. 5-fold training is used, applied to all three network's training. With 5-fold our model achieved $0.90 \pm 0.01$ Dice. Results reported by other works are present in Table~\ref{tab:harp_res}. Our methodology has similar performance to what is reported by Atalaglou et al.'s recent, simultaneous work~\cite{ataloglou2019fast}. Interestingly, the initial methodology of both methods is similar, in the use of multiple 2D CNNs.

\begin{table}[ht]
\centering
\begin{tabular}{cc}
\toprule
\textbf{Deep learning Methods} & \textbf{HarP (DICE)} \\\midrule
3D U-Net - Isensee et al.~\cite{isensee2017brain} (2017)& 0.86 \\
Hippodeep - Thyerau et al.~\cite{thyreau2018segmentation} (2018)& 0.85\\
QuickNat - Roy et al.~\cite{roy2019quicknat} (2019)& 0.80\\
\textbf{Ataloglou et al.~\cite{ataloglou2019fast} (2019)}& \textbf{0.90}* \\
\textbf{E2DHipseg (this work)} & \textbf{0.90}* \\\midrule
\textbf{Atlas-based methods}  & \\\midrule
FreeSurfer v6.0 \cite{fischl2012freesurfer} (2012)& 0.70\\
Chincarini et al. \cite{chincarini2016integrating} (2016) & 0.85\\
Platero et al. \cite{platero2017combining} (2017) & 0.85\\\bottomrule
\end{tabular}
  \caption{Reported testing results for HarP. This work is named E2DHipseg. Results with * were calculated following a 5-fold cross validation.}
  \label{tab:harp_res}
\end{table}

\subsubsection{HCUnicamp}

As described previously, the HCUnicamp dataset has lack of one of the hippocampi in many of it's scans (Figure~\ref{fig:bigdata}), and it was used to examine the generalization capability of these methods. Table~\ref{tab:nat_res} has mean and standard deviation Dice for all HCUnicamp volumes, using both masks, or only one the left or right mask, with multiple methods. ``with Aug.'' refers to the use of augmentations in training. We also report Precision and Recall, per voxel classification, where positives are hippocampus voxels and negatives are non hippocampus voxels. Precision is defined by $TP / (TP + FP)$ and Recall is defined by $TP / (TP + FN)$, where TP is true positives, FP are false positives and FN are false negatives. All tests were run locally. Unfortunately, we were not able to reproduce Atalaglou et al.'s method for local testing.

\begin{table}[ht]
\centering
\resizebox{\columnwidth}{!}{
\begin{tabular}{cccccc}
\toprule
\multicolumn{6}{c}{\textbf{HCUnicamp (Controls)}} \\\midrule
\textbf{Method} & \textbf{Both (Dice)} & \textbf{Left (Dice)} & \textbf{Right (Dice)} & \textbf{Precision} & \textbf{Recall}  \\\midrule
3D U-Net - Isensee et al.~\cite{isensee2017brain} (2017)& $0.80\pm0.04$ & $0.81\pm0.04$ & $0.78\pm0.04$ & $0.76\pm0.10$ & $0.85\pm0.06$ \\
Hippodeep - Thyerau et al.~\cite{thyreau2018segmentation} (2018) & $0.80\pm0.05$ & $0.81\pm0.05$ & $0.80\pm0.05$ & $0.72\pm0.10$ & $0.92\pm0.04$  \\
QuickNat - Roy et al.~\cite{roy2019quicknat} (2019)&  $0.80\pm0.05$  & $0.80\pm0.05$ & $0.79\pm0.05$ & $0.71\pm0.11$ & $\mathbf{0.92\pm0.04}$\\
\textbf{E2DHipseg without Aug.}  & $0.82\pm0.03$ & $0.83\pm0.03$ & $0.82\pm0.03$ & $\mathbf{0.78\pm0.10}$ & $0.88\pm0.06$  \\
\textbf{E2DHipseg with Aug.} & $\mathbf{0.82\pm0.03}$ & $\mathbf{0.83\pm0.03}$ & $\mathbf{0.82\pm0.04}$ & $0.78\pm0.10$ & $0.89\pm0.06$\\\midrule
\multicolumn{6}{c}{\textbf{HCUnicamp (Patients)}}  \\\midrule
3D U-Net - Isensee et al.~\cite{isensee2017brain} (2017)&$0.74\pm0.08$ & $0.48\pm0.39$ & $0.56\pm0.36$ & $0.66\pm0.12$ & $0.87\pm0.07$ \\
Hippodeep - Thyerau et al.~\cite{thyreau2018segmentation} (2018) & $0.74\pm0.08$ & $0.48\pm0.39$ & $0.57\pm0.37$ & $0.63\pm0.12$ & $0.91\pm0.06$\\
QuickNat - Roy et al.~\cite{roy2019quicknat} (2019)&$0.71\pm0.08$  & $0.47\pm0.38$ & $0.56\pm0.36$ &  $0.59\pm0.12$ & $\mathbf{0.92\pm0.06}$ \\
\textbf{E2DHipseg without Aug.}  & $\mathbf{0.77\pm0.07}$ & $0.49\pm0.40$ & $0.58\pm0.37$ & $\mathbf{0.69\pm0.11}$ & $0.88\pm0.07$ \\
\textbf{E2DHipseg with Aug.} & $0.76\pm0.07$ & $\mathbf{0.50\pm0.40}$ & $\mathbf{0.58\pm0.37}$ & $0.68\pm0.11$ & $0.89\pm0.07$ \\\bottomrule
\end{tabular}
}
\caption{Locally executed testing results for HCUnicamp. All 190 volumes from the dataset are included, and no model saw it on training. The 3D U-Net here is using the same weights from table~\ref{tab:harp_res}. QuickNat performs whole brain multitask segmentation, not only hippocampus.}
\label{tab:nat_res}
\end{table}

Our method performed better than other recent methods on the literature in the HCUnicamp dataset, even though HCUnicamp is not involved on our methodology development. However, no method was able to achieve more than 0.8 mean Dice in epilepsy patients. The high number of false positives due to hippocampus removal is notable by the low left and right DICE, and low precision. The impact of additional augmentations was not statistically significant in the epilepsy domain.

Our method takes around 15 seconds on a mid-range GPU and 3 minutes on a consumer CPU to run, per volume. All the code used on its development is available in \url{github.com/MICLab-Unicamp/e2dhipseg}, with instructions for how to run it in an input volume, under MIT license. A free executable version for medical research use, without enviroment dependencies, is available on the repository. To avoid problems with different head orientations, there is an option to use  MNI152 registration when predicting in a given volume. Even when performing registration, the output mask will be in the input volume's space, using the inverse transform. In regards to pre-processing requirements, our method requires only for the volume to be a 3D MRI in the correct orientation. The automatic MNI152 registration option solves this problem, in a similar way to Hippodeep. A GPU is recommended for faster prediction but not necessary.

\subsection{Adaptation to HCUnicamp}

Additional experiments were performed now involving HCUnicamp data in training, to try and learn to recognize the resection. The experiments involved making a hold-out separation of HCUnicamp. In the previous experiment, all volumes were involved in the testing and not used for training of any method. In this one, hold-out with 70\% training, 10\% validation and 20\% testing is performed with balance between control and patients, to allow for training. Note that these results are not comparable with other method's results or even or own results present in Table~{\ref{tab:nat_res}}, since the dataset is different and we are now training on part of HCUnicamp. To avoid confusion, the hold-out subsets will be refered to as HCU-Train and HCU-Test. Experiments were also performed including only control volumes or only patient volumes, with the same hold-out approach~(Table~{\ref{tab:hcadapt}}). Results improve when training on HCUnicamp volumes, but the high standard deviation still shows that the method is failing to recognize resections. 

\begin{table}[ht]
\centering
\begin{tabular}{cccc}
\toprule
\textbf{Trained on}   & \textbf{Both (Dice)}         & \textbf{Left (Dice)}         & \textbf{Right (Dice)}        \\\midrule
Patients              & $0.84\pm0.04$ & $0.60\pm0.41$ & $0.56\pm0.42$ \\
Controls and Patients & $0.86\pm0.05$ & $0.71\pm0.36$ & $0.74\pm0.34$ \\
Controls              & $0.90\pm0.01$ & $0.89\pm0.02$ & $0.90\pm0.01$ \\\bottomrule
\end{tabular}
\caption{E2DHipseg with networks trained in HCU-Train. Training is performed in all volumes, only patients or only controls, and testing is done in HCU-Test, also selecting for patients, controls, or all.}
\label{tab:hcadapt}
\end{table}

Another experiment attempts to learn from both datasets at the same time (Table~{\ref{tab:hcconcat}}). The dataset now is the concatenation of HarP and HCUnicamp. The datasets where mixed together with a 70\% training, 10\% validation and 20\% testing hold-out. The presence of patients and controls is balanced between the sets. Also included are results from testing in a different domain while training in other.

\begin{table}[ht]
\centering
\resizebox{\columnwidth}{!}{%
\begin{tabular}{ccccc}
\toprule
\textbf{Trained on} & \textbf{Tested on} & \textbf{Both (Dice)} & \textbf{Left (Dice)}         & \textbf{Right (Dice)}        \\\midrule
Harp-Train             & HCU-Test & $0.79\pm0.07$ & $0.65\pm0.33$ & $0.68\pm0.31$ \\
HCU-Train        & HarP-Test      & $0.50\pm0.29$ & $0.50\pm0.31$ & $0.50\pm0.29$ \\
Harp-Train + HCU-Train & HarP-Test      & $0.89\pm0.01$ & $0.89\pm0.01$ & $0.89\pm0.02$ \\
Harp-Train + HCU-Train & HCU-Test & $0.85\pm0.04$ & $0.69\pm0.35$ & $0.73\pm0.33$ \\\bottomrule
\end{tabular}%
}
\caption{This table compares training in one dataset and testing in the other. Betters results are achieved when involving both domains in training.}
  \label{tab:hcconcat}
\end{table}

E2DHipseg was able to achieve good Dice in both the HarP and HCU when both are involved on training. However, while looking at only left or right results, poor Dice standard deviation is still present, meaning problems with resection are still happening. While examining predictions from training only in HCU and testing in HarP, in many cases the method predicted a resection was present in darker scans, when it wasn't, resulting in high false negatives.

\subsection{Qualitative Results}
\begin{figure}[ht]
\begin{center}
\begin{minipage}[b]{\textwidth}
  \includegraphics[width=\textwidth]{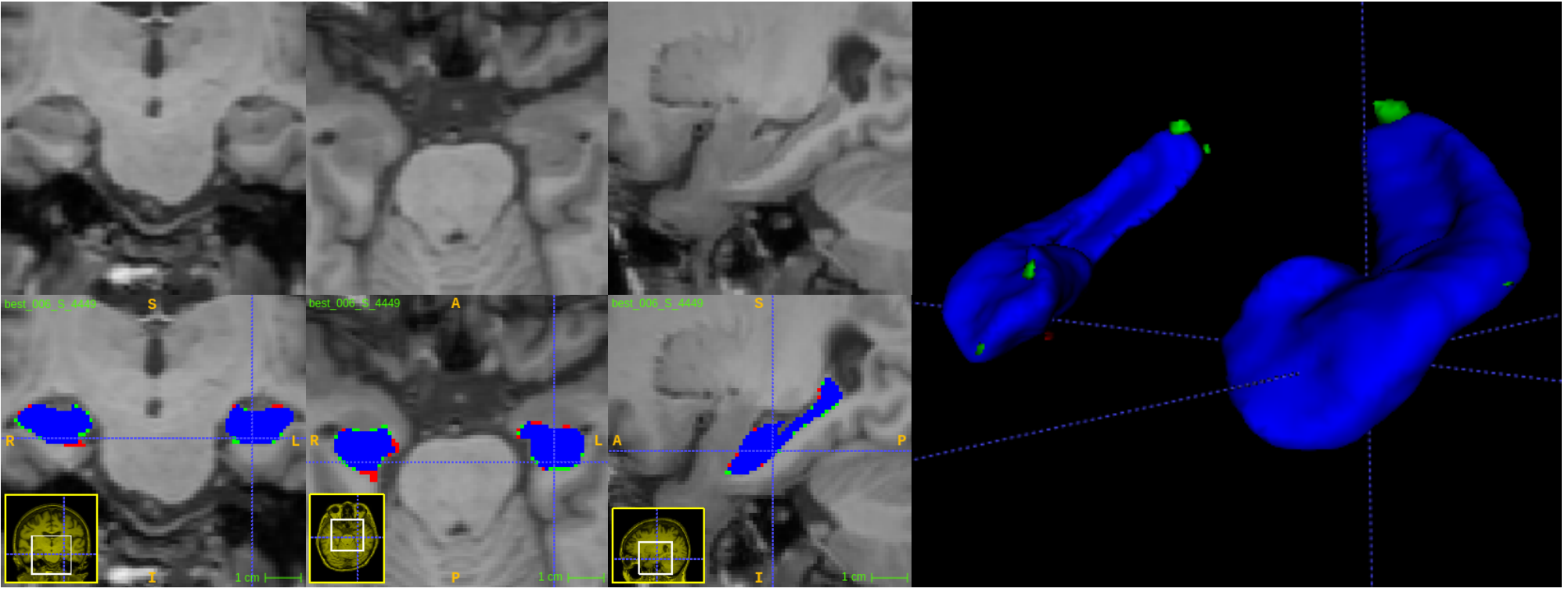}
  \centerline{(a)}\medskip
\end{minipage}
\begin{minipage}[b]{\textwidth}
  \includegraphics[width=\textwidth]{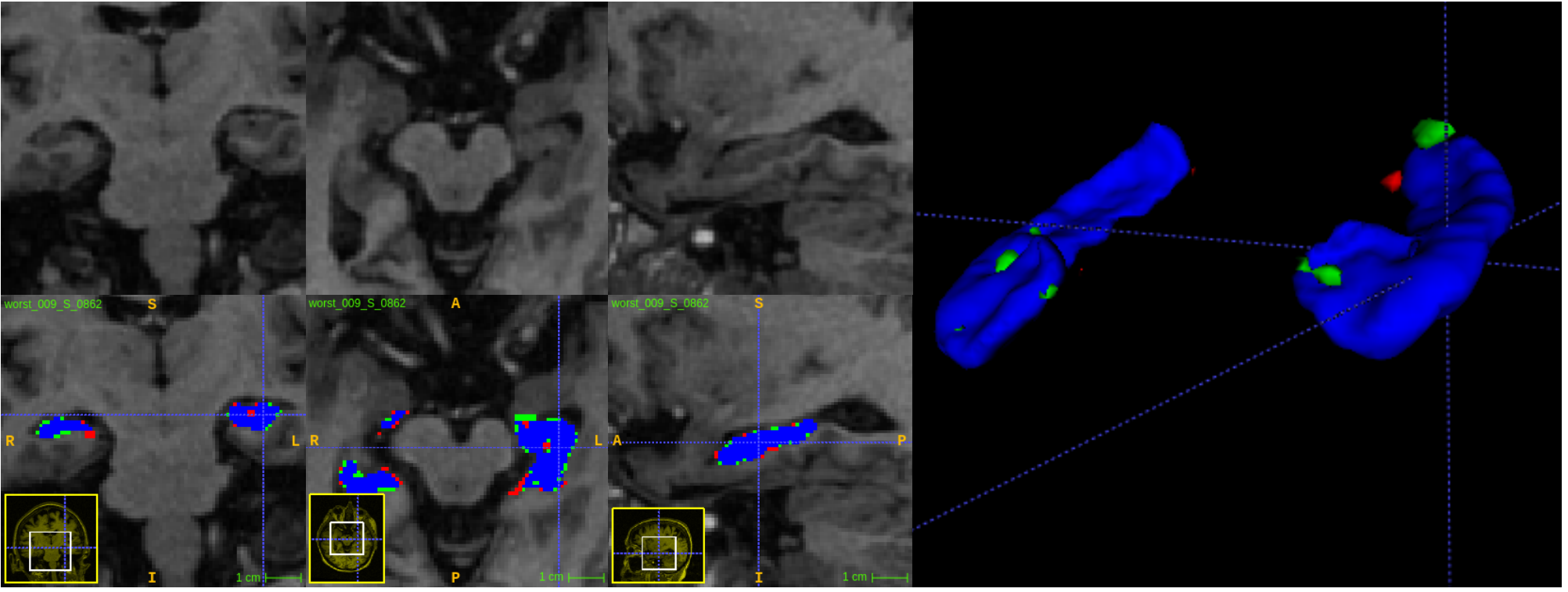}
  \centerline{(b)}\medskip
\end{minipage}

\caption{Multiview and 3D render (approximate) of our (a) best and (b) worst cases while evaluatin in the HarP test set. Prediction in green, target in red and overlap in purple.}
\label{fig:best_worst}
\end{center}
\end{figure}

\begin{figure}[ht]
\begin{center}
\begin{minipage}[b]{\textwidth}
  \includegraphics[width=\textwidth]{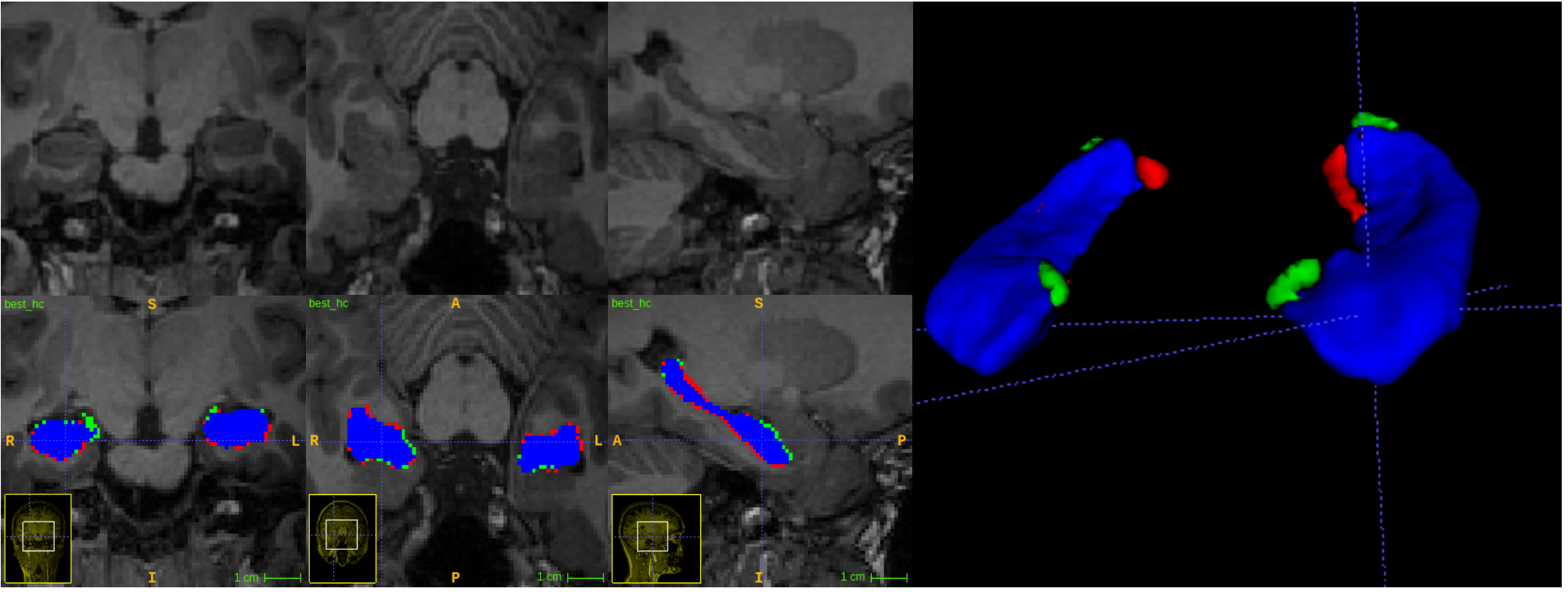}
  \centerline{(a)}\medskip
\end{minipage}
\begin{minipage}[b]{\textwidth}
  \includegraphics[width=\textwidth]{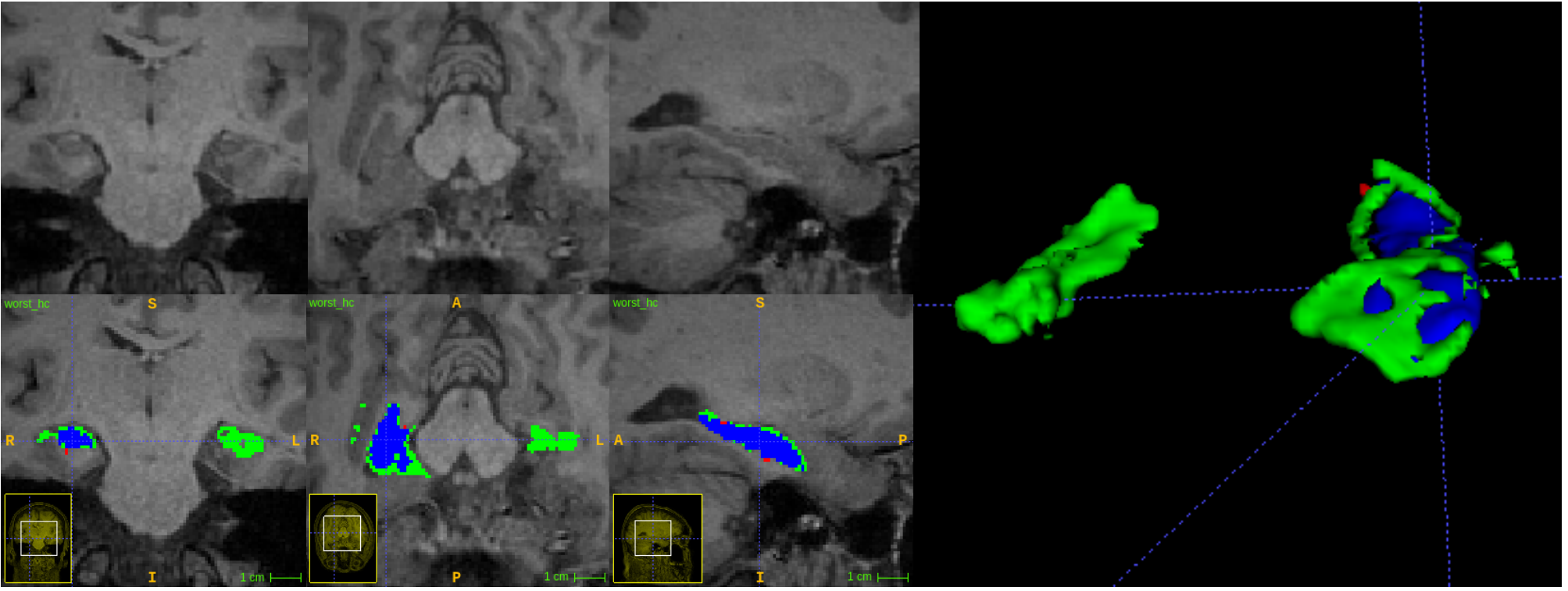}
  \centerline{(b)}\medskip
\end{minipage}

\caption{Multiview and 3D render of our (a) best and (b) worst cases while testing in the HCUnicamp dataset. Prediction in green, target in red and overlap in purple.}
\label{fig:best_worst_hc}
\end{center}
\end{figure}

While visually inspecting HarP results, very low variance was found. We noted no presence of heavy outliers. Other methods present similar, stable results.  

However, in HCUnicamp, way more errors are visible in the worst segmentations in Figure~\ref{fig:best_worst_hc}(b). Specially where the hippocampus is removed. Other methods have similar results, with false positives in voxels where the hippocampus would be in a healthy subject or Alzheimer's patient. As expected, the best segmentation, displayed in Figure~\ref{fig:best_worst_hc}(a), was in a control, healthy subject.


\section{Discussion}
\label{sec:discussion}

Regarding the Consensus approach from our method, most of the false positives some of the networks produce are eliminated by the averaging of activations followed by thresholding and post processing. This approach allows the methodology to focus on good segmentation on the hippocampus area, without worrying with small false positives in other areas of the brain. It was also observed that in some cases, one of the networks fails and the consensus of the other two ``saves" the result.

The fact that patches are randomly selected and augmented in runtime  means they are mostly not repeated in different epochs. This is different to making a large dataset of pre-processed patches with augmentation. We believe this random variation during training is very important to ensure the network keeps seeing different data in different epochs, improving generalization. This idea is similar to the Dropout technique~\cite{srivastava2014dropout}, only done in data instead of weights. Even with this patch randomness, re-runs of the same experiment resulted mostly in the same final results, within 0.01 mean Dice of each other.

As visible on the results of multiple methods, Dice when evaluating using the HCUnicamp dataset is not on the same level as what is seen on the public benchmark. Most methods have false positives on the removed hippocampus area, in a similar fashion to Figure~\ref{fig:best_worst_hc}(b). The fact that QuickNat and Hippodeep have separate outputs for left and right hippocampus does not seem to be enough to solve this problem. We believe the high false positive rate is due to textures similar to the hippocampus, present in the hippocampus area, after its removal.

Final experiments attempt to adapt the methodology to Epilepsy volumes. Training in HCUnicamp improved results, but the high standard deviation and mistakes on hippocampus resections are still present. A similar story is seen while analysing results from concatenating the HarP train and HCU-Train dataset in training. The method was able to achieve good overall Dice in both the HarP test set and HCU-Test, of 0.89 and 0.85, but while analysing right and left hippocampus separately the high standard deviation due to missed resections was still present. The resulting mean Dice was low due to cases of false positives in resections on the left or right Dice resulting in 0 Dice, pulling the mean Dice down drastically. This was confirmed in the qualitative results and does not happen when training and testing in HCUnicamp controls or HarP, as showcased by the similar, low standard deviation between overall Dice and left/right Dice. This problem could possibly be solved with a preliminary hippocampus presence detection phase in future work, but this is not in the scope of this paper, since HCUnicamp was used here as a test set and this approach would be a bias to the test set.

\section{Conclusion}
\label{sec:conclusion}
This paper presents a hippocampus segmentation method including consensus of multiple U-Net based CNNs and traditional post-processing, successfully using a new optimizer and loss function from the literature. The presented method achieves state-of-the-art performance on the public HarP hippocampus segmentation benchmark. The hypothesis was raised that current automatic hippocampus segmentation methods, including our own, would not have the same performance on our in-house epilepsy dataset, with many cases of hippocampus removal. Quantitative and qualitative results show failure from those methods to take into account hippocampus removal, in unseen epilepsy data. This raises the concern that current automatic hippocampus segmentation methods are not ready to deal with hippocampus resection due to epilepsy treatment. We show that training in the epilepsy data does improve results, but there is still room for improvement. In future work, improvements can be made to our methodology to detect the removal of the hippocampus as a pre-processing step.

\section*{Acknowledgements}
Alzheimer's disease data collection and sharing for this project was provided by the Alzheimer's Disease Neuroimaging Initiative
(ADNI) (National Institutes of Health Grant U01 AG024904) and DOD ADNI (Department of Defense award
number W81XWH-12-2-0012). ADNI is funded by the National Institute on Aging, the National Institute of
Biomedical Imaging and Bioengineering, and through generous contributions from the following: AbbVie,
Alzheimer’s Association; Alzheimer’s Drug Discovery Foundation; Araclon Biotech; BioClinica, Inc.; Biogen;
Bristol-Myers Squibb Company; CereSpir, Inc.; Cogstate; Eisai Inc.; Elan Pharmaceuticals, Inc.; Eli Lilly and
Company; EuroImmun; F. Hoffmann-La Roche Ltd and its affiliated company Genentech, Inc.; Fujirebio; GE
Healthcare; IXICO Ltd.; Janssen Alzheimer Immunotherapy Research \& Development, LLC.; Johnson \&
Johnson Pharmaceutical Research \& Development LLC.; Lumosity; Lundbeck; Merck \& Co., Inc.; Meso
Scale Diagnostics, LLC.; NeuroRx Research; Neurotrack Technologies; Novartis Pharmaceuticals
Corporation; Pfizer Inc.; Piramal Imaging; Servier; Takeda Pharmaceutical Company; and Transition
Therapeutics. The Canadian Institutes of Health Research is providing funds to support ADNI clinical sites
in Canada. Private sector contributions are facilitated by the Foundation for the National Institutes of Health
(www.fnih.org). The grantee organization is the Northern California Institute for Research and Education,
and the study is coordinated by the Alzheimer’s Therapeutic Research Institute at the University of Southern
California. ADNI data are disseminated by the Laboratory for Neuro Imaging at the University of Southern
California. 

Additionally, we thank our partners at BRAINN for letting us use their epilepsy dataset on this research.

Finally, we thank São Paulo Research Foundation (FAPESP) and CAPES for funding this research under grant 2018/00186-0 and CNPq research funding, process numbers 310828/2018-0
and 308311/2016-7. 

\bibliography{paper}

\pagebreak

\end{document}


\section*{Appendix}

\section*{Training}

\label{appendix}

This appendix presents some experiments related to optimizing our methodology as a whole, including choice of optimizer, loss functions, and the consensus approach.

\begin{figure}[ht]
\begin{center}
\begin{minipage}[b]{.49\textwidth}
  \includegraphics[width=\textwidth]{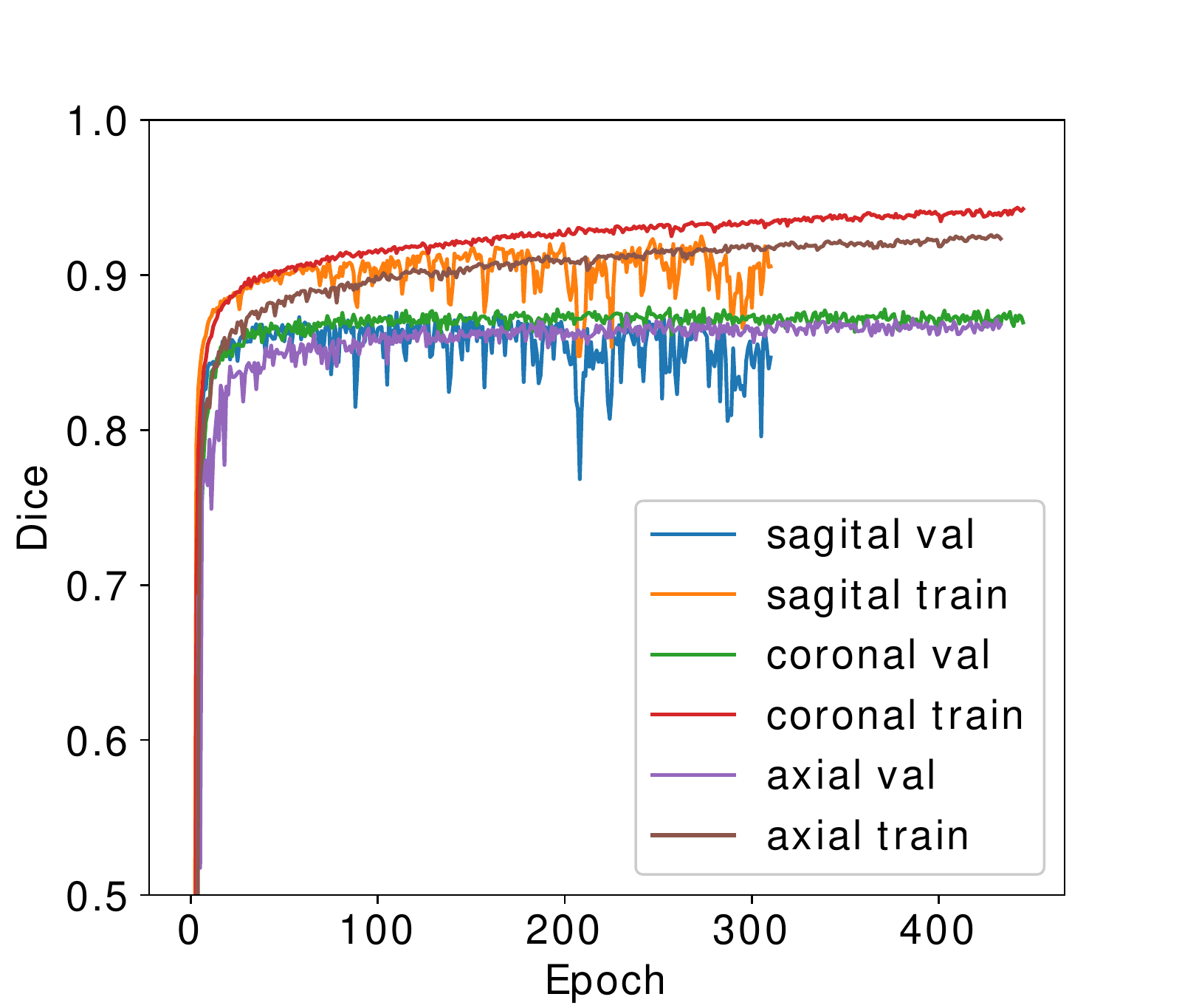}
  \centerline{(a)}\medskip
\end{minipage}
\begin{minipage}[b]{.49\textwidth}
  \includegraphics[width=\textwidth]{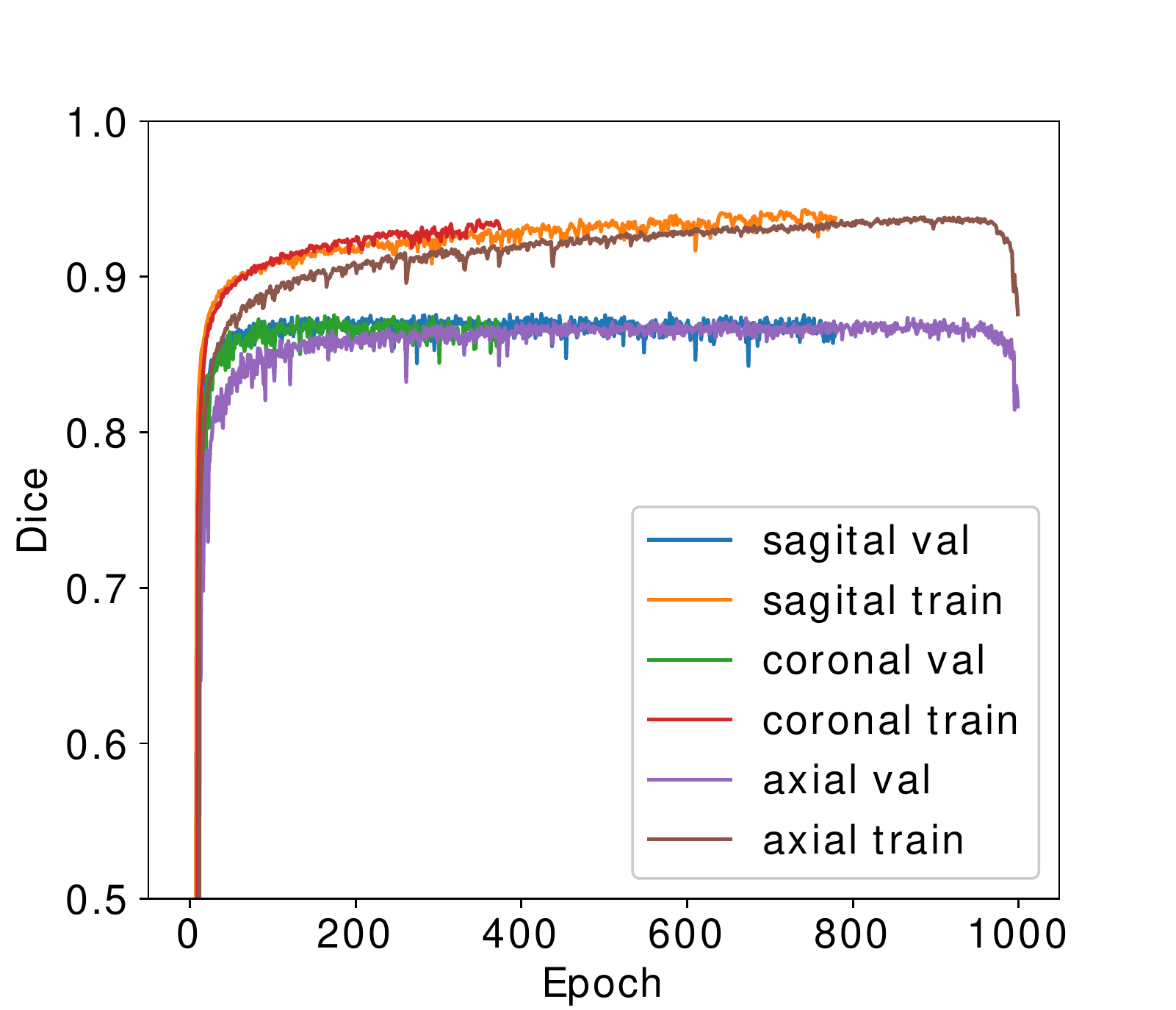}
  \centerline{(b)}\medskip
\end{minipage}

\caption{Validation and training Dice for all models, using: (a) ADAM (b) RADAM. Both with same hyperparameters and no stepping. Early stopping is due to patience. RADAM displays more stability. Although overfit is noticeable in (b), training was stopped by patience and only the best validation weights are used.}
\label{fig:trains}
\end{center}
\end{figure}

\subsection*{Optimizers, Learning Rate and Scheduling}
Training hyperparameters are the same for all three networks. Regarding the optimizer of choice and initial LR, grid search defined 0.0001 with ADAM~\cite{kingma2014adam} and 0.005 LR with SGD~\cite{bengio2015deep} to deliver similar performance. The recent RADAM from Liu et al.~\cite{liu2019variance} with 0.001 initial LR ended up being the optimizer of choice, due to improved training stability and results (Fig S.~\ref{fig:trains}). LR reduction scheduling is used, with multiplication by 0.1 after 250 epochs, its impact is showcased on Figure S.~\ref{fig:best_train}(a). While training on HarP with an 80\% holdout training set, an epoch consisted of going through around 5000 sagittal, 4000 coronal and 3000 axial random patches extracted from slices with presence of hippocampus, depending on which network is being trained, with a batch size of 200. The max number of Epochs allowed is 1000, with a patience early stopping of no validation improvement of 200 epochs. Note that weights are only saved for the best validation Dice.

\subsection*{Hyperparameter Experiments}
Some of the most important hyperparameter experiments can be seen in Table S.~\ref{tab:hyper}. These showcase the impact of Boundary Loss and RAdam in relation to more traditional approaches. Results from each change in methodology were calculated using the full consensus and post-processing. For these experiments, holdout of 80/20\% on HarP was used, keeping Alzheimer's labels balanced. Reported Dice is the mean over the 20\% test set.

\begin{table}[ht]
\centering
\begin{tabular}{ccccccc}
\toprule
\textbf{Optimizer} & \textbf{LR} &\textbf{Loss} & \textbf{HarP (Dice)} \\\midrule
SGD                & 0.005               & Dice Loss    & 0.8748 \\
ADAM               & 0.0001              & Dice Loss    & 0.8809        \\
ADAM               & 0.0001              & GDL           & 0.8862 \\
ADAM               & 0.0001              & Boundary      & 0.9068        \\
RADAM              & 0.0001              & Boundary      & 0.9071  \\
RADAM              & 0.001               & Boundary       & \textbf{0.9133}  \\
\bottomrule
\end{tabular}
\caption{Some of the most relevant hyperparameters experiments test results, in a hold-out approach to HarP. The bolded result represents the final model. All tests in this table use $64^{2}$ patch size and the modified U-Net architecture.}
  \label{tab:hyper}
\end{table}

\begin{figure}[ht]
\begin{center}
\begin{minipage}[b]{.42\textwidth}
  \includegraphics[width=\textwidth]{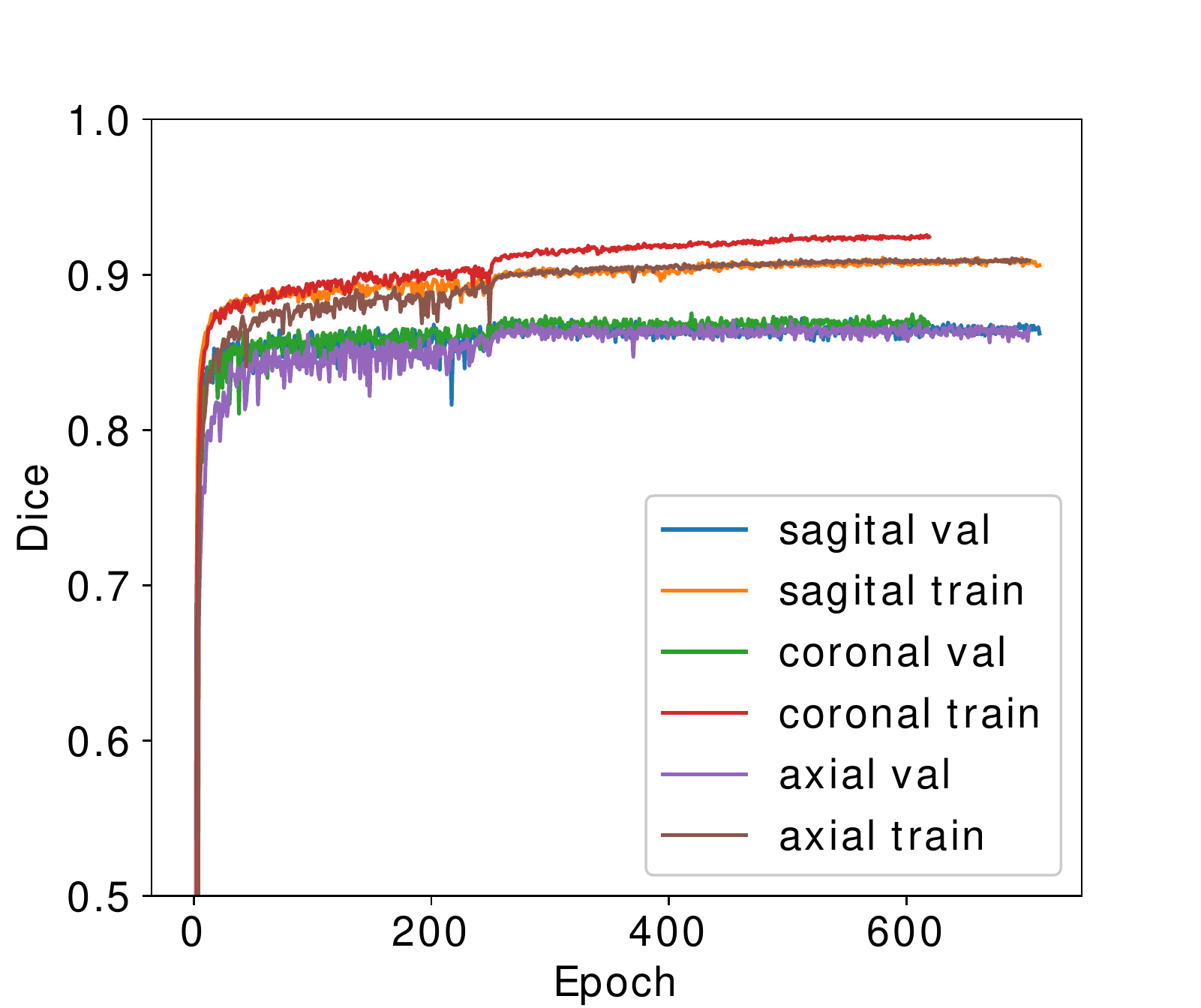}
  \centerline{(a)}\medskip
\end{minipage}
\begin{minipage}[b]{.49\textwidth}
  \includegraphics[width=\textwidth]{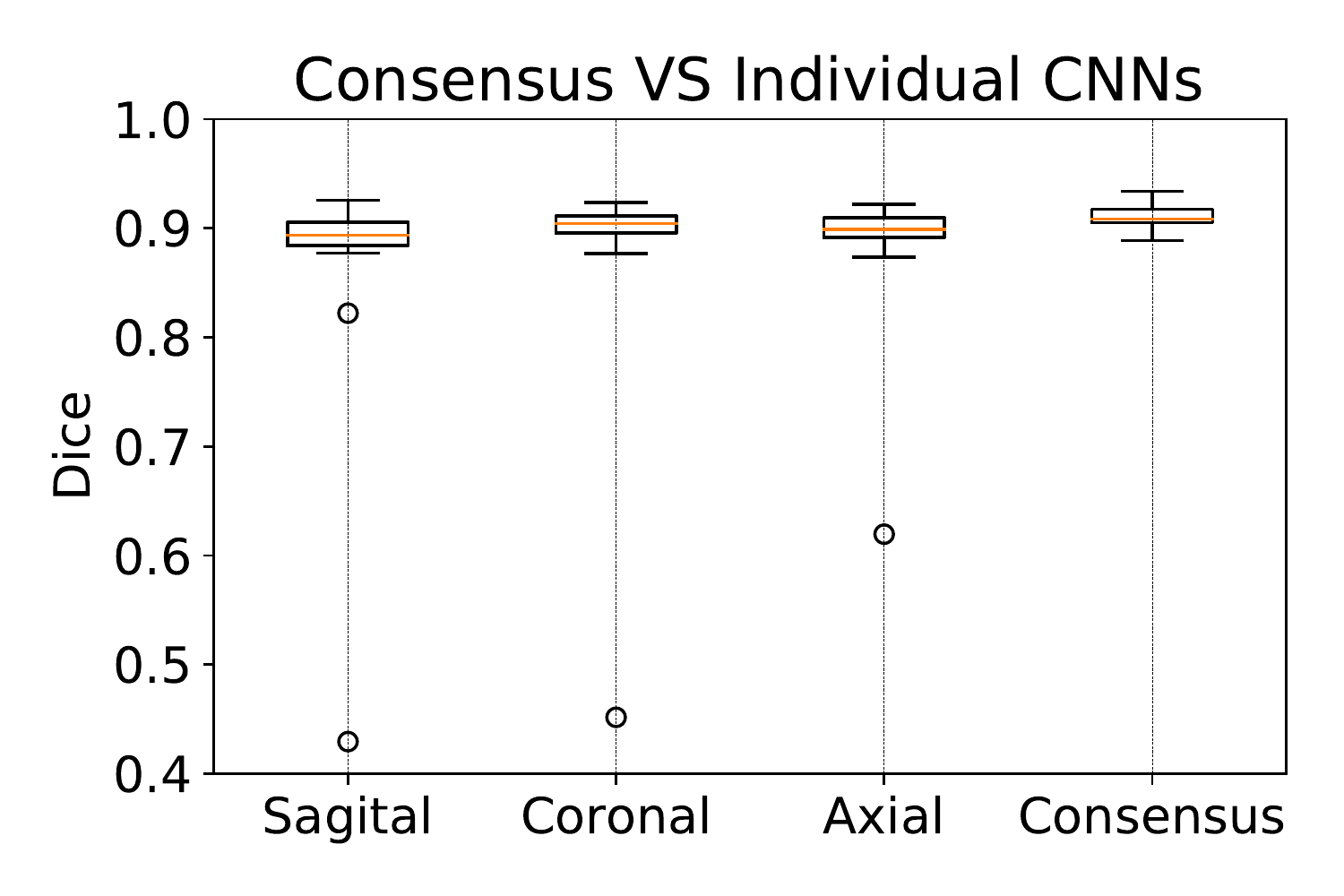}
  \centerline{(b)}\medskip
\end{minipage}

\caption{(a) Training and validation Dice curve for the best model, with RADAM and LR step. (b) Boxplot for HarP test models, showing the improvement in variance and mean Dice from the Consensus compared to using only one network.}
\label{fig:best_train}
\end{center}
\end{figure}

Early experiments showed that for the patch selection strategy, 80/20\% provided the best balance between positive and negative patches, with $64^{2}$ patch size. Implementation of Boundary Loss resulted in slightly better test Dice than Dice Loss. We found that augmentation techniques only impacted Dice results in HarP slightly, sometimes even making results worse. Augmentation's most relevant impact, however, was avoiding overfitting and very early stopping due to no validation improvements in some cases, leading to unstable networks.

We found that, as empirically expected, the consensus of the results from the three networks brings less variance to the final Dice as seen in Figure S.~\ref{fig:best_train}(b), where the result of isolated networks are evaluated in comparison to the consensus. Early studies confirmed that 0.5 is a reasonable value to choose for threshold after the activation averaging. Attempts at using a fourth 3D UNet as a consensus generator/error correction phase did not change results significantly. Since the best performing network varied according to hyperparameters, we choose to keep a simple average of activations instead of giving more weight to one of the networks.

\bibliography{paper}